\documentclass[%
 reprint,
 showpacs,preprintnumbers,
 amsmath,amssymb,
 aps,
 prb,
]{revtex4-1}
\usepackage{dcolumn}
\usepackage{subfigure}
\usepackage{here}
\usepackage{graphicx,color}
\usepackage{mathrsfs}

\makeatletter
\def\btt#1{\texttt{\@backslashchar#1}}
\DeclareRobustCommand\bblash{\btt{\@backslashchar}} \makeatother

\begin{document}

\title{Spin-dependent refraction at the interface of lateral heterostructures of 2$H$-type transition-metal dichalcogenide monolayers}

\author{Tetsuro Habe}
\affiliation{Department of Applied Physics, Hokkaido University, Sapporo, Hokkaido 060-0808, Japan}

\date{\today}

\begin{abstract}
We study the refraction effect of electronic wave in hole-doped lateral heterojunctions of metallic and semiconducting transition-metal dichalcogenide monolayers.
This effect is theoretically investigated in 2$H$-type MoSe$_2$-NbS$_2$ and WSe$_2$-NbS$_2$ junctions by combining the first-principles calculation and the lattice Green's function method.
We show that the electronic waves change the direction of motion at the interface and collimate the velocity along two different directions depending on the spin.
We find that the transmission probability increases with the charge density and that the direction of refracted electron beams is close to $\pm30^\circ$ with respect to the perpendicular axis to the interface.
The metallic transition-metal dichalcogenide is essential for the refraction effect because of the strong trigonal-warping effect, the large Fermi surface, and the Zeeman-type spin-orbit coupling.
The refraction effect enables to generate the spin-polarized electronic current by using a simple fabrication of transition-metal dichalcogenide monolayers.
\end{abstract}

\maketitle
\section{Introduction}
Transition-metal dichalcogenides(TMDCs) are atomic-layered materials in which atomically thin crystals are stacked by van der Waals interaction.
The 2$H$-type monolayers host two-dimensional electronic system with the strong spin-orbit coupling.\cite{Xiao2012,Zibouche2014}
The electronic states are characterized by the valley degree of freedom; two valleys in semiconducting TMDCs\cite{Xiao2012} and three valleys in metallic TMDCs\cite{Habe2019-1}, where they split into two spin states due to the Zeeman-type spin-orbit coupling depending on the valley.
The correlation between the spin and valley degrees of freedom leads to novel phenomena and applications in spintronics\cite{Shan2013,Klinovaja2013,Suzuki2014,Yuan2014,Habe2015,Shao2016,Habe2017,Kormanyos2018,Ciccarino2018,Berman2019} and valleytronics\cite{Cao2012,Zeng2012,Gong2013,Ye2016,Lundt2019}.

In Ref.\ \onlinecite{Habe2015}, the author and coworker proposed the spin-dependent refraction of electronic waves in hole-doped TMDCs with an atomic step, i.e., the junction of TMDC monolayer and bilayer.
Conducting electrons change the direction of motion at the interface of two regions similar to the boundary of different refraction indexes for light.
Since the direction depends on the spin, we suggested that the monolayer-bilayer junction works as a spin splitter. 
However, the refracted angle is also depending on the wave number along the interface and thus the refracted electron beams are not well collimated.
In some directions of motion, electrons have both up-spin and down-spin.
Thus the complete separation of two spins is not impossible except the junctions with a slight transmission probability.
In the current work, we consider the lateral heterojunction of metallic and semiconducting TMDC monolayers and reveal that the lateral heterojunction enables to completely separate two spins with a high transmission probability and to generate highly-collimated and spin-polarized electron beams.

The lateral heterojunction is composed of different atomic layers which are atomically bounded as a single layer. 
Experimentally, such junctions have been realized by using chemical vapor deposition (CVD) on a substrate.\cite{Huang2014,Gong2014,Duan2014,Chen2015,Chen2015-2,Zhang2015,He2016,Xie2018}
The junction of semiconducting TMDCs has been theoretically and experimentally investigated and applied to the p-n junction,\cite{Li2015,Najmzadeh2016}, valleytronics,\cite{Ullah2017} and optelectronics\cite{Son2016}. 
However, the metal-semiconductor junction of TMDCs does not attract attentions comparing with that of semiconducting TMDCs.\cite{Habe2019-2}
This paper exhibits the significant advantage of the metal-semiconductor TMDC junction in spintronics.

In this paper,  we theoretically investigate the electronic transport in the hole-doped lateral heterojunction of metallic and semiconducting TMDC monolayers by combining the first-principles calculation and the lattice Green's function method.
We adopt MoSe$_2$ or WSe$_2$ as the semiconducting monolayer and NbS$_2$ as the metallic monolayer because the lattice constants are nearly equivalent among these materials.

\begin{figure}[htbp]
\begin{center}
 \includegraphics[width=65mm]{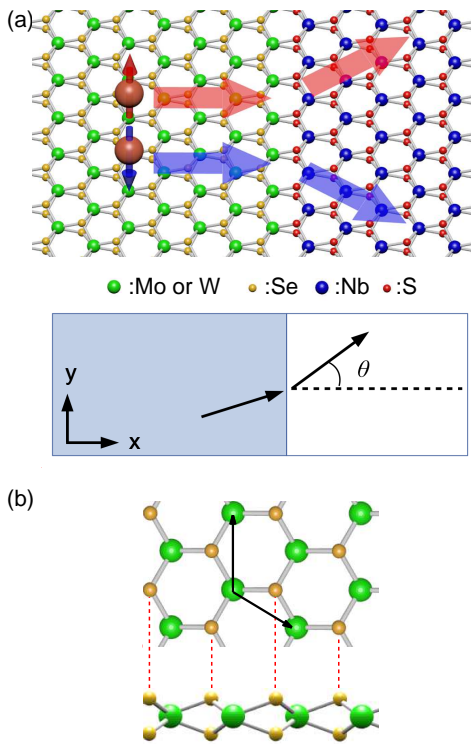}
\caption{
The schematics of spin-dependent refraction of electron in (a). The lower panel depicts the upper junction and indicates the definition of angle for the direction of motion. In (b), the crystal structure is shown in the vertical view (upper) and the horizontal view (lower). The solid arrows represent the primitive lattice vectors.
 }\label{fig_schematic}
\end{center}
\end{figure}
\section{Calculation method}
We numerically calculate the transmission probability in the lateral heterojunction of semiconducting and metallic TMDC monolayers with the zig-zag interface and simulate the difference of velocity between the incident wave and the transmitted wave.
The electronic states in the junction are described by a multi-orbital tight-binding model where the hopping integrals and the on-site potentials are calculated from the first-principles bands.
The transmission probability is numerically obtained by using the lattice Green's function method.

\subsection{First-principles band calculation}
We obtain the band structures of pristine TMDC monolayers to construct the multi-orbital tight-binding model describing electronic states in lateral heterojunction of the semiconducting and metallic monolayers. 
The 2$H$-type structure is adopted as the crystal structure; one sublayer of transition-metal atoms is sandwiched between two sublayers of chalcogen atoms and a two-dimensional hexagonal lattice is formed as shown in Fig.\ \ref{fig_schematic}.
We consider MoSe$_2$ and WSe$_2$ as the material for the semiconducting region, and NbS$_2$ as that for the metallic region.
The electronic band structures are calculated by using quantum-ESPRESSO,\cite{Quantum-espresso} a numerical code based on the density functional theory (DFT), in the projector augment-wave (PAW) method with considering the spin orbit interaction.
The cut-off energy of plane wave basis 50 Ry and the convergence criterion 10$^{-8}$ Ry are adopted as numerical parameters.
The lattice constants of these materials are also adopted the relaxed crystal structure simulated by using the same code.
In the tree materials, the distance between two sublayers of calcogen atoms is different; $d_{\mathrm{Se}-\mathrm{Se}}=$3.343\AA\ for MoSe$_2$, 3.360\AA\ for WSe$_2$, and $d_{\mathrm{S}-\mathrm{S}}=3.127$\AA\ for NbS$_2$.
The lattice constant of in-plane honeycomb lattice is almost equivalent to each other with the mismatch less than about 1\%; 3.319\AA\ for MoSe$_2$, 3.317 for WSe$_2$, and 3.346 for NbS$_2$ according the first-principle calculation.

\begin{figure}[htbp]
\begin{center}
 \includegraphics[width=65mm]{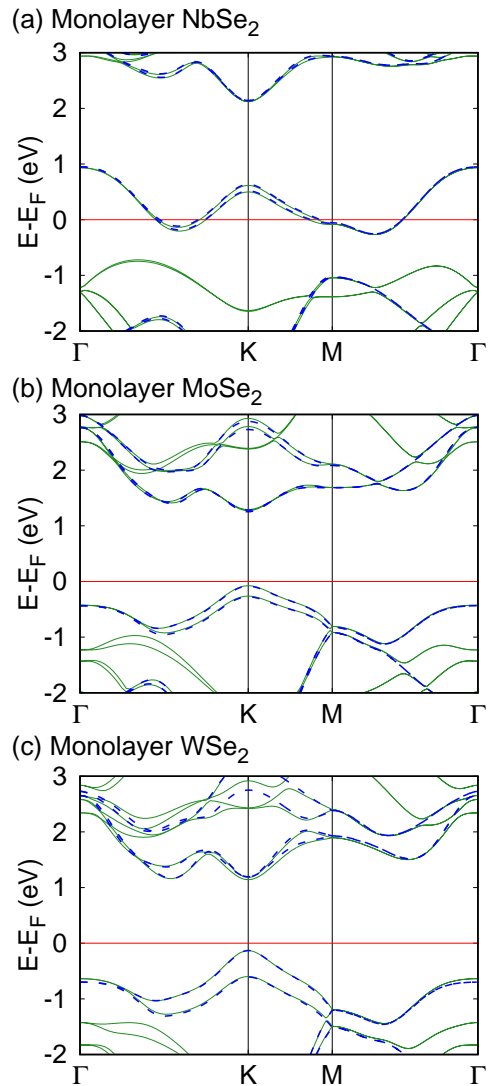}
\caption{
The band structures of (a) NbS$_2$, (b) MoSe$_2$, and (c) WSe$_2$ monolayers. The solid and dashed lines indicate the first-principles bands and the bands calculated by using the multi-orbitals tight-binding model. The horizontal line represent the Fermi level for the pristine crystal.
 }\label{fig_band_structure}
\end{center}
\end{figure}
\begin{figure*}[htbp]
\begin{center}
 \includegraphics[width=160mm]{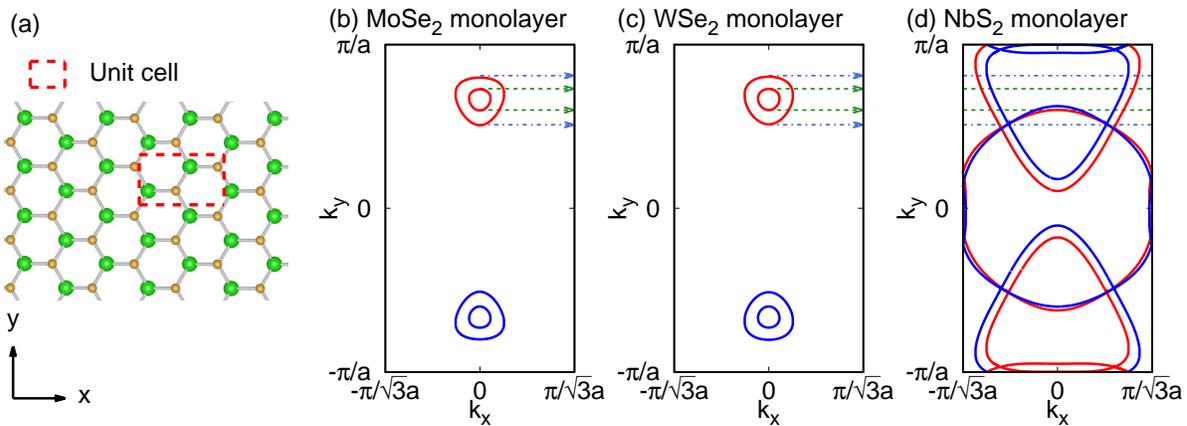}
\caption{
The crystal structure and the Fermi surface of TMDC. In (a), the unit cell is represented by the dashed line on the crystal structure. In (b), (c), and (d), the Fermi surface of  MoSe$_2$, WSe$_2$, and NbS$_2$ are shown, respectively. The charge density is $n=1\times10^{13}$cm$^{-2}$ for the smaller pockets and $n=5\times10^{13}$ cm$^{-2}$ for the larger pockets in (b) and (c), and $n=0$ in (d). The red and blue lines indicate the up-spin and down-spin states, respectively.
 }\label{fig_Fermi_surface}
\end{center}
\end{figure*}
We show the band structures of pristine MoSe$_2$, WSe$_2$, and NbS$_2$ monolayers in Fig.\ \ref{fig_band_structure}.
In general, the DFT calculation underestimates the band gap of semiconductor.\cite{Mak2010,Shi2013}
Thus we use the charge density instead of the Fermi energy as a parameter.
The charge density per unit cell is defined as
\begin{align}
n=n_0-\int^{\infty}_{-\infty}\frac{d^2\boldsymbol{k}}{(2\pi)^2}\sum_{n}\theta(E_F-E_{n\boldsymbol{k}}),
\end{align}
where $\theta(x)$ and $E_{n\boldsymbol{k}}$ are the step function and the energy dispersion of $n$-th band in Fig.\ \ref{fig_band_structure}, respectively.
Here $n_0$ represents the positive charge density due to the nuclei in the unit cell.
In the monolayer system, the total charge density can be controlled by using the gate.
In the following calculation, we consider a homogeneous gate to dope charges into the heterojunction and assume a uniform charge density far from the interface.
The condition can be achieved by adopting the Fermi level calculated from pristine monolayer as that in the each region of lateral heterojunction.
Around the interface, on the other hand, the band bending leads to the fluctuation of charge density.

\subsection{Representation by tight-binding model}
We construct a multi-orbitals tight-binding model to describe electronic states in the junction and assume that the hopping integrals and the on-site potentials to be those in the pristine monolayers.
The hopping integral and on-site potential are calculated by using Wannier90,\cite{Wannier90} a code to compute the maximally localized Wannier functions and these parameters from first-principles bands.
The atomic orbitals can be classified into two groups by the parity for the mirror operation along the out-of-plane direction.
The electronic bands around the Fermi level consist of the orbitals with even parity, three $d$-orbitals; $|d_{3z^2-r^2}\rangle$, $|d_{xy}\rangle$, and $|d_{x^2-y^2}\rangle$, and three $p$-orbitals; $|p_x^+\rangle$, $|p_y^+\rangle$, and $|p_z^-\rangle$, where $|p_\mu^\pm\rangle$ is the superposition of two $p_\mu$-orbitals in upper and lower chalcogen atoms.
The tight-binding Hamiltonian reproduces the first-principles bands as shown in Fig.\ \ref{fig_band_structure}.
At the interface, the hopping matrix for NbS$_2$ is adopted in this calculation where the hopping matrix for MoSe$_2$ adds 5\% to the transmission probability.

The transmission probability in the lateral hetero-junction is calculated by using the lattice Green's function method.
We adopt the unit cell as shown in Fig.\ \ref{fig_Fermi_surface}(a) for preparing the tight-binding model only with the nearest neighbor hopping in the $x$-direction as,
\begin{align}
H=\sum_{j,k_y}\left(\hat{c}_{k_y,j}^\dagger \hat{h}_{k_y,j}\hat{c}_{k_y,j}+\left[\hat{c}^\dagger_{k_y,j+1}\hat{t}_{k_y,j}\hat{c}_{k_y,j}+\mathrm{h.c.}\right]\right),
\end{align}
where$\hat{c}_{k_y,j}$ is the vector of annihilation operators for orbitals in the unit cell at $j$.
Here $h_{k_y,j}$ and $t_{k_y,j}$ describe the intra-cell hopping and inter-cell hopping, respectively. 
In the heterojunction, the wave number $k_y$ is preserved because the interface is commensurate in the $y$-direction due to the match of lattice constant between two monolayers.
The hopping matrices, $\hat{h}_{n}(k_y)$ and $\hat{t}_{n+1,n}(k_y)$, are those in MoSe$_2$ or WSe$_2$ for $n\leq0$ and in NbS$_2$ for $1\leq n$.
In the lattice Green's function method, the junction is separated into three regions; the left lead, the scattering region, and the right lead, where the hopping matrix is unchanged in the two leads.
We consider the electrons are coming from the left lead and transmit to the right lead, i.e., the left and right leads are the source and drain leads, respectively.

The incident and transmitted electronic waves are represented by the product of the vector component $\hat{c}_{k_y}$ and the position-dependent phase factor $\lambda^m$ at $m$.
In the TMDC monolayers, the spin-orbit coupling splits the up-spin and the down-spin in the perpendicular direction to the layer due to the mirror symmetry along the direction.
Thus the incident and transmitted waves can be classified into the two spin states as $\hat{c}_{k_y,m}=(\hat{c}_{\uparrow,k_y,m},\hat{c}_{\downarrow,k_y,m})$.
They can be obtained as the eigenvalue and the eivenvector of the equation describing the translation,
\begin{align}
\lambda\begin{pmatrix}
\hat{c}_{s,k_y,m}\\
\hat{c}_{s,k_y,m-1}
\end{pmatrix}=
\begin{pmatrix}
\hat{t}^{-1}_{k_y}(E_F-\hat{h}_{k_y})&\hat{t}^{-1}_{k_y} \hat{t}^\dagger_{k_y}\\
1&0
\end{pmatrix}
\begin{pmatrix}
\hat{c}_{s,k_y,m}\\
\hat{c}_{s,k_y,m-1}
\end{pmatrix},
\end{align}
with the Fermi energy $E_F$, where $\hat{h}_{k_y}$ and $\hat{t}_{k_y}$ are the hopping matrices in each monolayer.
The electrons are traveling in the channels with $|\lambda|=1$, which are extended states in the leads, and transmit between electronic states in the left and right leads with preserving $k_y$.
The direction of electronic motion can be represented by the velocity $(v_x,v_y)$ which is calculated by the expectation value of the velocity operator,
\begin{align}
\hat{v}_x=2i\begin{pmatrix}
0&-\hat{t}^\dagger_{k_y}\\
\hat{t}_{k_y}&0
\end{pmatrix},
\end{align}
with $(\hat{c}_{k_y,m},\hat{c}_{k_y,m-1})$, and 
\begin{align}
\hat{v}_y=\frac{1}{i\hbar}\frac{\partial}{\partial k_y}H
\end{align}
with $\sqrt{2}\hat{c}_{k_y,m}$.

\subsection{Lattice Green's function method}
The transmission probability is calculated by using the lattice Green's function method.\cite{Ando1991}
We give a brief review for this method below.
In this formulation, the self-energies for each $k_y$ in the left and right leads are given by
\begin{align}
\Sigma_{L,0}=t_{k_y,R}^\dagger F_{L,-}^{-1},\;\;\Sigma_{R,M+1}=t_{k_y,R}F_{R,+},
\end{align}
where the subscript, $L$ and $R$, indicates the left and right leads, respectively, and $x=0$ ($x=M+1$) indicates the boundary of the left(right) lead.
Here $F_{\pm}$ consists of the phase matrix $\Lambda_{\pm}=\mathrm{diag}[\lambda^{(1)}_\pm,\cdots,\lambda^{(N)}_\pm]$ and the matrix of basis $U_\pm=[\hat{c}^{(1)}_{k_y,\pm},\cdots,\hat{c}^{(N)}_{k_y,\pm}]$ as
\begin{align}
F_{\pm}=U_{\pm}\Lambda_{\pm}U_{\pm}^{-1},
\end{align}
where the electric states with the subscript $+(-)$ are right(left)-going waves, which decay or have the velocity in the positive(negative) $x$-direction.
%The Green's function in each lead is calculated as 
%\begin{align}
%G_{L(R)}(k_y,E)=(E-\hat{h}_{k_y,L(R)}-\Sigma_{L(R)}(k_y))^{-1},
%\end{align}
%at the boundary.
The on-site Green's function in the scattering region is given by
\begin{align}
G_{j,j}(k_y,E)=(E-\hat{h}_{k_y,j}-\Sigma_{L,j}-\Sigma_{R,j})^{-1},
\end{align}
with
\begin{align}
\Sigma_{L,j}=&\hat{t}_{k_y,j-1}^\dagger G_{L,j-1}(k_y,E)\hat{t}_{k_y,j-1},\\
\Sigma_{R,j}=&\hat{t}_{k_y,j} G_{R,j+1}(k_y,E)\hat{t}_{k_y,j}^\dagger.
\end{align}
Here $G_{L(R),j}(k_y,E)$ can be obtained by an iterative calculation,
\begin{align}
G_{L(R),j}(k_y,E)=(E-\hat{h}_{k_y,j}-\Sigma_{L(R),j})^{-1}.
\end{align}
Then the propagation from the left lead to a site $j$ is described by the lattice Green's function,
\begin{align}
G_{(j,0)}(k_y,E)=G_{j,j}(k_y,E)\hat{t}^\dagger_{k_y,j}G_{R,(j-1,0)}(k_y,E),
\end{align}
with the right-going Green's function,
\begin{align}
G_{R,(l,0)}(k_y,E)=G_{l}(k_y,E)\hat{t}^\dagger_{k_y,l}G_{R,(l-1,0)}(k_y,E),
\end{align}
from $G_{R,(0,0)}(k_y,E)=G_{R,0}(k_y,E)$.
The transmission coefficient is given by 
\begin{align}
T_{\alpha\beta}(k_y,E)=\sqrt{\frac{v_{x,\alpha}}{v_{x,\beta}}}&\{U_{R,+}^{-1}G_{M+1,0}(k_y,E)\hat{t}_{k_y,L}^\dagger\nonumber\\
&\times[F_{L,+}^{-1}-F_{L,-}^{-1}]U_{L,-}\}_{\alpha,\beta},
\end{align}
from the $\beta$ channel in the left lead to the $\alpha$ channel in the right lead.

\begin{figure}[htbp]
\begin{center}
 \includegraphics[width=70mm]{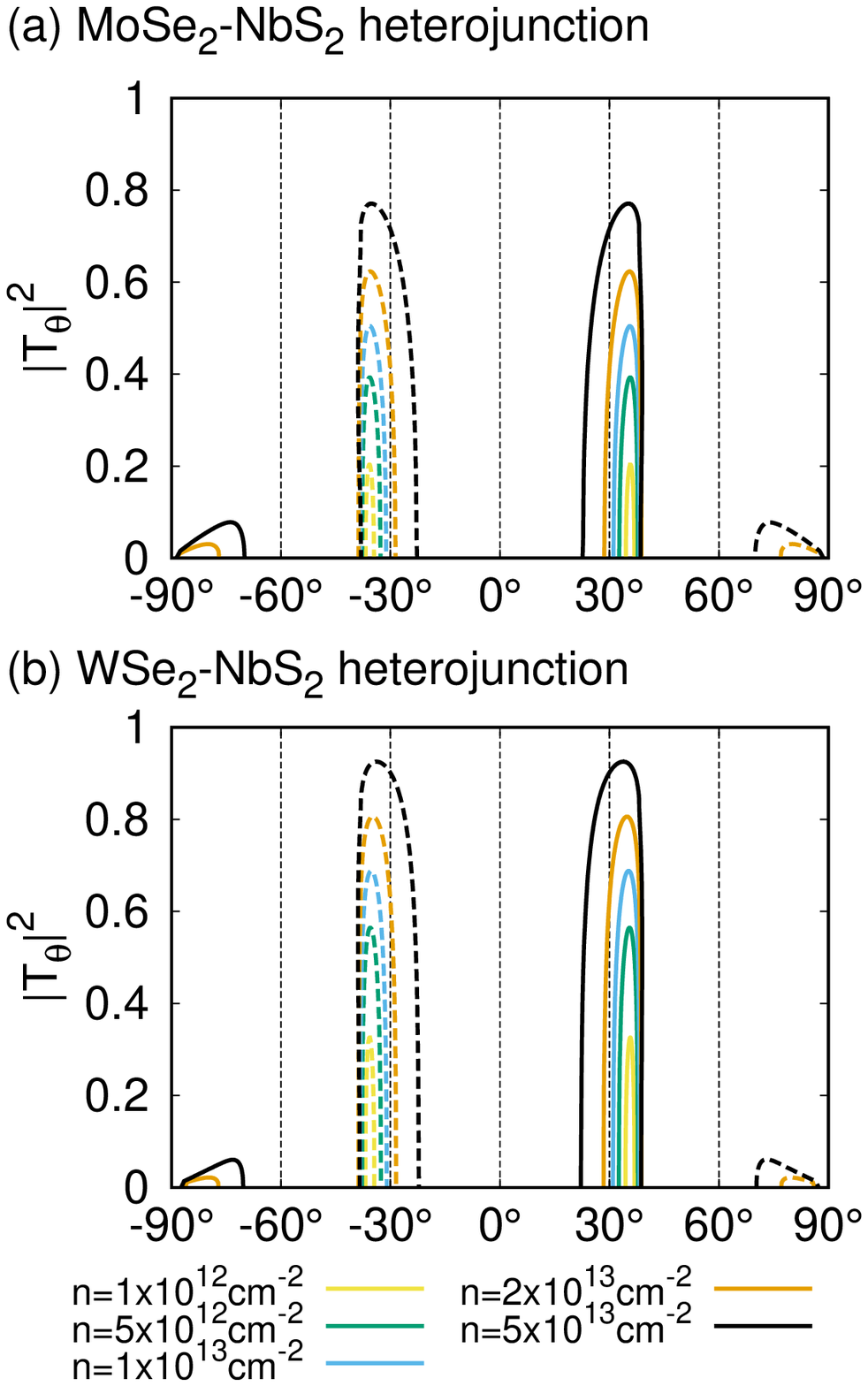}
\caption{
The transmission probability in the lateral heterojunction of MoSe$_2$ and NbS$_2$ in (a), and WSe$_2$ and NbS$_2$ in (b). The angle in the vertical axis indicates the direction of motion for transmitted electrons. The solid and dashed lines represent the transmission of up-spin and down-spin electrons, respectively.
 }\label{fig_transmission_probability}
\end{center}
\end{figure}
\section{Numerical results}
Firstly, we show the transmission probability for each transmitted electron in Fig.\ \ref{fig_transmission_probability}, where the angle in the vertical axis indicates the direction of motion of transmitted electron with respect to the $x$-axis.
We find up(down)-spin electrons transmit with a large probability along the direction of 30$^\circ$(-30$^\circ$) with respect to the $x$-axis.
The transmission probability increases with the charge density in both MoSe$_2$-NbS$_2$ and WS$_2$-NbS$_2$ heterojunctions but the degradation of convergence occurs  to the transmitted electrons.d
Moreover, the electrons transmit along the direction of $\pm80^\circ$ above $n=2\times10^{13}$ cm$^{-2}$.
However, the most probable direction of motion remains in 30$^\circ$.

The collimated flow of electrons is attributed to the trigonal-warping of the Fermi surface in NbS$_2$ monolayer.
We show the Fermi surface of MoSe$_2$, WSe$_2$, and NbS$_2$ in Fig.\ \ref{fig_Fermi_surface}, where we consider non-zero charge density in the semiconducting TMDCs; MoSe$_2$ and WSe$_2$.
In the leads, the electronic states coincide with those in pristine monolayer asymptotically.
Up(down)-spin electrons incident from the semiconducting layer propagate in the upper(lower) pockets.
Since electrons preserve $k_y$ in the transmission process, the transmitted electron changes its velocity in the NbS$_2$ region, where the electronic velocity is along the perpendicular direction to the Fermi surface.

In the NbS$_2$ monolayer, there are three Fermi pockets corresponding to three valleys.\cite{Habe2019-1}
The upper and lower pockets correspond to the K and K$'$ valleys and they are strongly deformed due to the trigonal warping effect.
In Fig.\ \ref{fig_Fermi_surface}, we represent the momentum region where conducting channels are present for up-spin electrons in both sides of heterojunction.
In this momentum region, the Fermi line of upper pocket are nearly linear.
Thus, the velocity is collimated in a direction, which is 30$^\circ$ with respect to the $x$-axis.
When the charge density is larger than $n=1\times10^{13}$cm$^{-2}$, electrons can transmit to the third pocket enclosing the $\Gamma$ point.
The transmission to channels in the $\Gamma$ valley produces the electrons with the velocity along -80$^\circ$ in Fig.\ \ref{fig_transmission_probability}.

\begin{figure}[htbp]
\begin{center}
 \includegraphics[width=70mm]{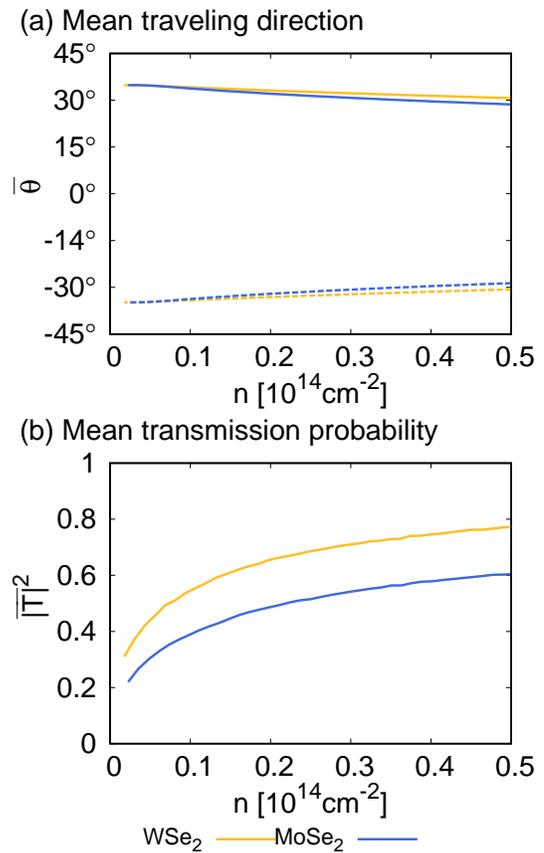}
\caption{
The mean value of traveling direction (a) and transmission probability (b) for transmitted electrons. In (a), the solid and dashed lines represent up-spin and down-spin electrons, respectively.
 }\label{fig_mean_value}
\end{center}
\end{figure}
Next, we discuss the average behavior of transmitted electrons in NbS$_2$ region.
In Fig.\ \ref{fig_mean_value}(a), the channel-averaged direction of motion is shown as a function of charge density $n$.
The mean value is defined by $\bar{\theta}=\arctan(\bar{v}_y/\bar{v}_x)$ with the mean velocity,
\begin{align}
\bar{v}_\mu=\frac{1}{N_y}\sum_{k_y,\alpha,\beta}v_{\mu,\alpha}|T_{\alpha\beta(k_y,E_F)}|^2,
\end{align}
where $N_y$ and $v_{\mu,\alpha}$ are the number of conducting channels in the semiconducting region and the velocity of electronic state $\alpha$ in the metellic region, respectively.
The up-spin and down-spin electrons propagate in the different directions on average.
The direction approaches $\pm30^\circ$ at $n=0.5\times10^{-14}$cm$^{-2}$.
Moreover, we consider the channel-averaged transmission probability defined by
\begin{align}
\bar{|T|}^2=\frac{1}{N_y}\sum_{k_y,\alpha,\beta}|T_{\alpha\beta(k_y,E_F)}|^2,
\end{align}
and give the probability  as a function of charge density in Fig.\ \ref{fig_mean_value}(b).
The numerical calculation reveals that WSe$_2$ provides a larger transmission probability than MoSe$_2$.
The transmission probability increases with the charge density in both junctions of MoSe$_2$ and WSe$_2$.

Finally, we compare these results with the refraction effect in semiconducting bilayer-monolayer junctions.
The electron beams are well collimated with a high transmission probability in the metal-semiconductor junction compared with the previous work about semiconducting monolayer-bilayer junctions in Ref.\ \cite{Habe2015}.
In the semiconducting bilayers, the electronic states are changed from those in monolayers by the inter-layer mixing of atomic orbitals and thus the transmission probability is suppressed.
Moreover, in the metallic monolayer, the Fermi surface is much larger and the trigonal warping effect is stronger than those in semiconducting materials.
As shown in Fig.\ \ref{fig_Fermi_surface}, the large and strongly trigonal-warping Fermi surface leads to the well-collimated electron beams.

\section{Conclusion}
We investigated the  spin-dependent refraction of electronic waves in the hole-doped lateral heterojunction of metallic and semiconducting TMDC monolayers with a zig-zag interface.
We consider two junctions of MoSe$_2$-NbS$_2$ and WSe$_2$-NbS$_2$ junctions where the mismatch of lattice constant is less than 1\% between two layers.
The transmitted waves are separated into two different directions depending on the spin degree of freedom.
The up-spin and down-spin electron beams are well collimated in $30^\circ$ and $-30^\circ$, respectively, with respect to the perpendicular axis to the interface.
In both the junctions, the transmission probability increases with the charge density.
We have shown that the spin-dependent refractio with a high transmission probability is attributed to the large and strongly trigonl-warping Fermi surface of metallic NbS$_2$ monolayer.
\bibliography{TMDC}
\end{document}